\documentclass[aps,prd,twocolumn,superscriptaddress,nofootinbib,showpacs]{revtex4}
\usepackage[pass,letterpaper]{geometry}
\pdfoutput=1  
\usepackage{graphicx,color}
\usepackage{latexsym,amsmath,amsfonts,amssymb,booktabs}
\usepackage[pdfencoding=auto]{hyperref}
\usepackage{slashed,upgreek,amscd,cancel,tensor,color}
\usepackage{url}
\usepackage{placeins}
\usepackage{doi}
\usepackage[normalem]{ulem}

\definecolor{MyBlue}{rgb}{0.15,0.15,0.70}
\hypersetup{
colorlinks=true,
citecolor=MyBlue,
linkcolor=MyBlue,
urlcolor=MyBlue
}

\newcommand{\be}{\begin{equation}}
\newcommand{\ee}{\end{equation}}
\newcommand{\beq}{\begin{equation}}
\newcommand{\eeq}{\end{equation}}
\newcommand{\bea}{\begin{eqnarray}}
\newcommand{\eea}{\end{eqnarray}}

\newcommand{\dd}{\text{d}}

\usepackage[stable]{footmisc}

\newcommand\ees{\end{eqnarray}}
\newcommand\bees{\begin{eqnarray}}

\begin{document}
\title{Comment on the article ``Anisotropies in the astrophysical gravitational-wave background:
The impact of black hole distributions'' by A.C. Jenkins {\it et al.} [{\tt arXiv:1810.13435}] }
\author{Giulia Cusin}
\email{giulia.cusin@physics.ox.ac.uk}
\affiliation{Astrophysics Department, University of Oxford, DWB, Keble Road, Oxford OX1 3RH, UK}
\author{Irina Dvorkin}
\email{irina.dvorkin@aei.mpg.de}
\affiliation{Max Planck Institute for Gravitational Physics (Albert Einstein Institute), Am M\"{u}hlenberg 1, Potsdam-Golm, 14476, Germany}
\author{Cyril Pitrou}
\email{ pitrou@iap.fr}
\affiliation{Institut d'Astrophysique de Paris, CNRS UMR 7095, \\
Institut Lagrange de Paris, 98 bis, Bd Arago, 75014 Paris, France}
\author{Jean-Philippe Uzan}
\email{uzan@iap.fr}
\affiliation{Institut d'Astrophysique de Paris, CNRS UMR 7095, \\
Institut Lagrange de Paris, 98 bis, Bd Arago, 75014 Paris, France}
\date{\today}

\begin{abstract}
We investigate the discrepancy pointed out by
Jenkins et al. in Ref.~\cite{Mairi2}  between the predictions of
anisotropies of the astrophysical gravitational wave (GW) background, derived using different methods in  Cusin et al.~\cite{Cusin:2018rsq} and in Jenkins et al.~\cite{Mairi1}. We show that
this discrepancy is not due to our treatment of galaxy clustering,
contrary to the claim made in Ref.~\cite{Mairi2} and we show that our
modeling of clustering gives results in very good agreement with
observations. Furthermore we detail that the power law spectrum used in Refs.~\cite{Mairi2} and \cite{Mairi1} to describe galaxy
clustering is incorrect on large scales and leads to a different scaling for the
multipoles $C_\ell$. Moreover, we also explain that the analytic derivation of the gravitational wave background correlation function in
Refs.~\cite{Mairi2} and \cite{Mairi1} is mathematically ill-defined and predicts an amplitude of the angular power spectrum which depends on the (arbitrary) choice of a non-physical cut-off. 
\end{abstract}
\maketitle

Following the development of a framework to describe the anisotropies of the stochastic gravitational wave (GW) background~\cite{Cusin:2017fwz,Cusin:2017mjm}, two predictions for its amplitude in the LIGO frequency band have been proposed in the literature: the first predictions were presented in our letter \cite{Cusin:2018rsq} followed by those of Ref.~\cite{Mairi1}. Both studies consider the contribution of binary black hole mergers and rely on the same analytic framework  designed in Ref.~\cite{Cusin:2017fwz}.

As explained in Refs.~\cite{Cusin:2017fwz}-\cite{Cusin:2017mjm}, the expression for anisotropies has two components entering in a multiplicative way, see e.g. Eq.~(72) of  Ref.~\cite{Cusin:2017fwz}: (1) an {\em astrophysical} part which describes the process of GW  emission inside a galaxy and which depends on the details of sub-galactic physics and (2) a {\em cosmological} component which describes the transfer function of cosmological perturbations, galaxy clustering and GW propagation along the line of sight. Refs.~\cite{Cusin:2018rsq} and~\cite{Mairi1} differ in at least two aspects: the choice of the astrophysical model and the treatment of density perturbations and galaxy clustering. For the latter, Ref.~\cite{Mairi1} relies both on an analytic approach and on the input from the Millennium Simulation while Ref.~\cite{Cusin:2018rsq} describes the clustering by means of a bias function while metric perturbations, velocities and matter overdensity are evolved from initial power spectra after inflation using a Bolzmann code (CMBquick \cite{CMBquick}) and including non linearities using Halofit~\cite{Smith:2002dz,Halofit2012}. The astrophysical models used in Refs.~\cite{Mairi1} and  \cite{Cusin:2018rsq} are also different. The resulting amplitude of the angular power spectrum differs between these two studies. On this basis, it was not clear whether this discrepancy is due to the different astrophysical model used in the two predictions, or to the different description of galaxy clustering.

This discrepancy was recently investigated in Ref.~\cite{Mairi2} by
some of the authors of Ref.~\cite{Mairi1}. They use the astrophysical
model proposed in Ref.~\cite{Mairi1} and then derive predictions for
the GW angular power spectrum both using the approach of
Ref.~\cite{Mairi1} and the one of  Ref.~\cite{Cusin:2018rsq} to
describe galaxy clustering. They find a discrepancy of two orders of
magnitude between these two calculations (see in particular Fig.~2 in
Ref.~\cite{Mairi2}) and conclude that its origin lies in the description of
galaxy clustering, since the astrophysical model was taken to be the
same. As a consequence, they claim that the method of
Ref.~\cite{Cusin:2018rsq} fails in describing galaxy clustering. We
explain in the following why this conclusion is erroneous. Since in 
Refs.~\cite{Mairi1, Mairi2} (Jenkins et al. hereafter) the analytic approach to the
description of clustering gives results claimed to be consistent with
the ones obtained from the numerical simulation, we focus in the
following on the analytic approach of Jenkins et al. We show that the
power law correlation function used by  Jenkins et al. does not provide a
realistic description of galaxy clustering. Moreover, we show that the
mathematical approach used in Jenkins et al. to compute the GW background
correlation function is not correct and leads to a result which
depends on the (arbitrary) choice of a nonphysical cut-off, introduced
to regularize an otherwise divergent integral. Summarizing, we show in
detail that the approach of Jenkins et al. used to compute the angular
power spectrum of the background leads to: (1) an incorrect estimate
of the shape of the angular power spectrum of the background (2) an
arbitrariness in the prediction of its amplitude, since it is cut-off dependent. We also explain how to obtain
consistent results, with a proper (and standard) analytic treatment of
the galaxy correlation function. Sticking to our astrophysical
model, we derive in a mathematically consistent way the angular power
spectrum of the GW background, for both the power-law galaxy correlation
function used by Jenkins et al. and for the more realistic one of
Ref.~\cite{Cusin:2018rsq}. In particular, our standard approach does
not suffer from mathematical inconsistencies. We then address the
issue of the slope of the spectrum and we show that the difference
between the slope of Ref.~\cite{Cusin:2018rsq} and the one that would
be obtained using the power law correlation function of Jenkins et
al. is due to an overestimation of the contributions from large scales. We use the recent cosmological parameters of Ref.~\cite{Planck2015} throughout.

 Jenkins et al. use a power law galaxy correlation function  (see Eq.~(64) of Ref.~\cite{Mairi2})
\be\label{Defxi}
\xi_{\rm Gal}(r) = \left(\frac{r}{d_1}\right)^{-\gamma}
\ee
with either a spectrum estimated from the VIMOS survey~\cite{VIMOS13}
with central values $d_1=4.29 h^{-1}{\rm Mpc}$ and $\gamma=1.63$ or
with parameters fitted to the numerical simulations whose central
values are $d_1=5.05 h^{-1}{\rm Mpc}$ and $\gamma=1.67$. It is claimed
in Jenkins et al., and we checked it as well, that the results are nearly not affected by these differences, hence we choose the VIMOS values here. We note that the VIMOS values quoted are describing the galaxy correlation function at an average redshift of $z = 0.97$. Hence in order to describe correctly the correlation function which is dominated by contributions at low redshift, it is necessary to correct it for its evolution, and as a simple estimate we consider here that the spectrum~(\ref{Defxi}) needs to be multiplied by the linear growth factor
\be\label{DefGrowth}
g^2 \equiv \left[\frac{D_+(z=0)}{D_+(z=0.97)}\right]^2 \simeq 1.63^2\simeq 2.64\,.
\ee
From now on, $\xi_{\rm Gal}$ refers to $g^2 \xi_{\rm Gal}$. 

\begin{figure}[!htb]
\includegraphics[width=\columnwidth]{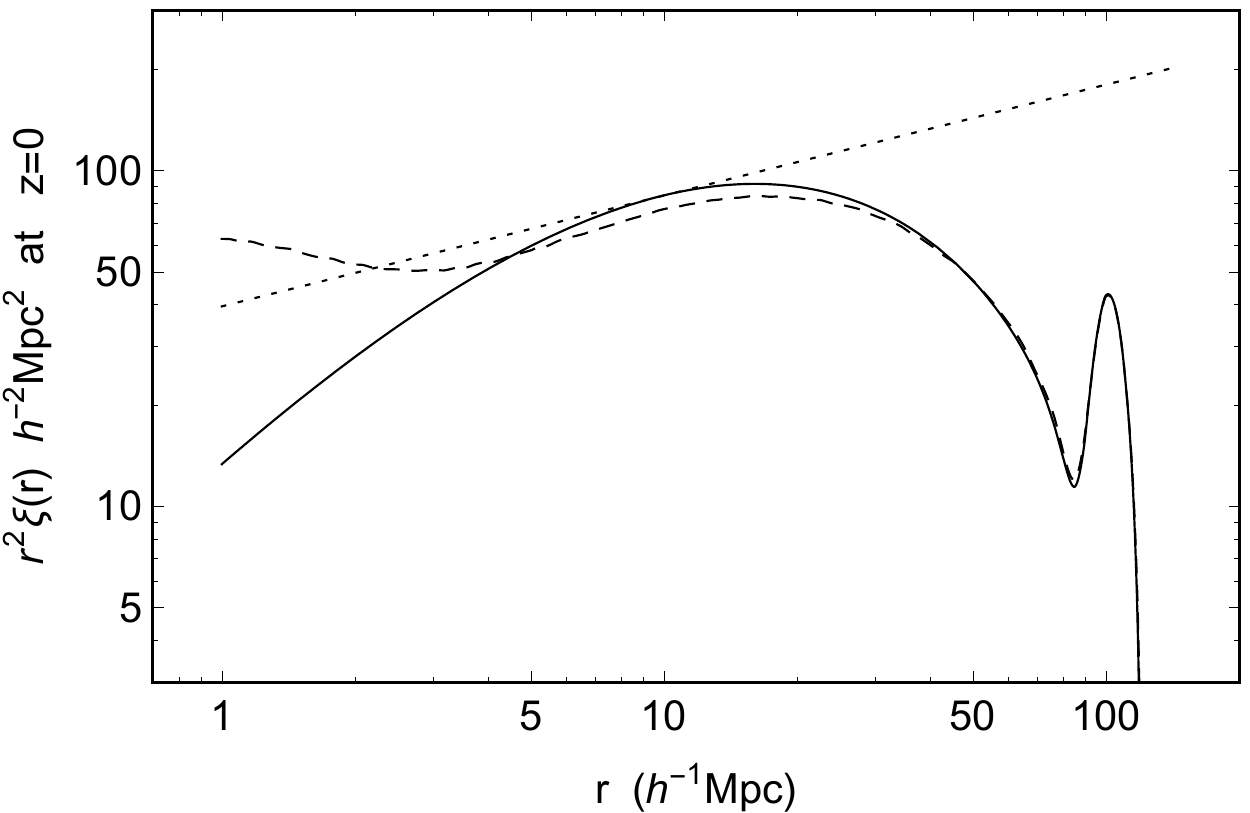}
\caption{Correlation function used in our treatment with and without
  Halofit (dashed and solid line respectively) and the analytic
  approximation used by Jenkins et al. (dotted line).}\label{schiaccio}
\end{figure}

\begin{figure}[!htb]
\includegraphics[width=\columnwidth]{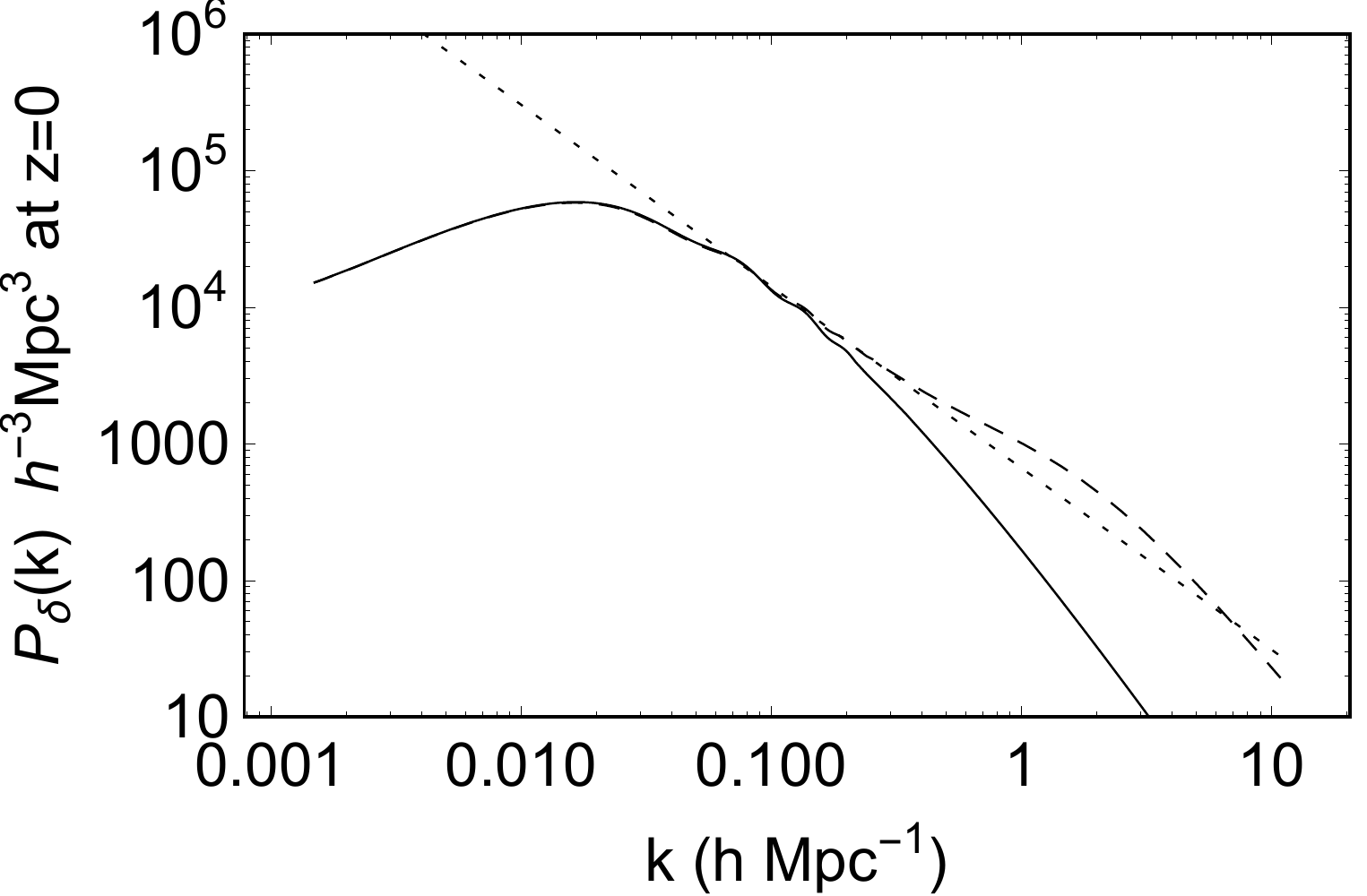}
\caption{Power spectrum used in our treatment with and without Halofit
  (dashed and solid line respectively) and the analytic approximation
  for the power spectrum derived from Jenkins et al. (dotted line).}\label{Pschiaccio}
\end{figure}

The galaxy power spectrum can be inferred from the galaxy correlation function using a simple Hankel transformation~\cite{Reimberg2015} as
\be\label{HankelxitoP}
P_{\rm Gal}(k) = 4\pi \int \xi_{\rm Gal}(r) j_0(kr)r^2 \dd r \,. 
\ee
When considering the power law correlation function (\ref{Defxi}), the power spectrum is also a power law and reads
\be\label{PGalk}
P_{\rm Gal}(k) = g^2 \frac{4\pi}{k^3} (k d_1)^\gamma \Gamma(2-\gamma)
\sin(\pi \gamma/2)\,.
\ee
Note that from the power spectrum, the correlation function is also inferred from a Hankel transformation as
\be\label{HankelPtoxi}
\xi_{\rm Gal}(r) = \int \frac{k^2 \dd k}{2 \pi^2} P_{\rm Gal}(k) j_0(kr)\,.
\ee
The galaxy correlation function and power spectrum of Jenkins et al. are plotted in dotted lines in Figs.~\ref{schiaccio} and \ref{Pschiaccio}.

In our approach~\cite{Cusin:2017fwz}, the power spectrum is obtained from the linear evolution of the nearly scale invariant power spectrum set by inflation, corrected by {\rm Halofit} to account for the late-time non-linearities, and applying a scale-invariant bias $b(z) = b_0\sqrt{1+z}$ relating the matter fluctuations to the galaxy fluctuations. Even though we do not use the correlation function in our developments, we deduce it here for comparison using Eq.~(\ref{HankelPtoxi}). Our galaxy correlation function and power spectrum are depicted in Figs.~\ref{schiaccio} and \ref{Pschiaccio}, in continuous line with only the linear evolution, and in dashed line with the Halofit contribution. Choosing $b_0=1.5$ leads to a very good agreement with the BOSS data, as can be seen by comparing the correlation function (with Halofit correction) to the Figs. 3 and 4 of Ref.~\cite{Anderson}, or the power spectrum (also with Halofit correction) to Fig. 8 of Ref.~\cite{Anderson} (the agreement is very good with the reported bias $b(z=0.57) \simeq 1.87$).

The comparison of the spectra and correlation functions of
Figs.~\ref{schiaccio} and \ref{Pschiaccio} shows that the simple power
law description of Jenkins et al. matches our description on the
scales $2 h^{-1}{\rm Mpc}\lesssim r \lesssim 20 h^{-1}{\rm Mpc}$. Most
notably, the power law correlation function overestimates scales
larger than $20 h^{-1}{\rm Mpc}$, a feature which happens to be
crucial for the analytic understanding as detailed below. In the power
spectrum comparison of Fig~\ref{Pschiaccio}, it is also apparent that
the power-law spectrum of Jenkins et al. fails to capture the smaller Fourier modes.

We have shown in Ref.~\cite{Cusin:2017fwz,Cusin:2017mjm} how the
multipoles of the fluctuations of the GW background from astrophysical
sources can be estimated. Given that the 2-point correlator of the
cosmological fluctuations is diagonal in Fourier space due to
statistical homogeneity, it is standard practice to derive theoretical
expressions for the angular correlation function of any cosmological
observable (or its corresponding $C_\ell$) which involve integrations
on the power spectrum. In particular this allows one to obtain
rigorously the plane parallel approximation (see
e.g. Ref.~\cite{Reimberg2015}), and the Limber approximation for
extended sources (see e.g. Refs.~\cite{LoVerde:2008re,PitrouFlat}).
However instead of relying on a Limber approximation for the
expression for the angular correlation function, Jenkins et al. assume that
the correlation function between a point in a direction $\hat{e}_1$ and
at redshift $z_1$ (that is at the associated comoving distance $r_1(z_1)$
on our past line cone) and another point in direction $\hat{e}_2$ and
at redshift $z_2$, is given by
\be\label{WrongLimber}
\langle \delta_{\rm Gal}(z_1,\hat{e}_1) \delta_{\rm Gal}(z_2,\hat{e}_2)\rangle
= \xi_{\rm Gal}(r_{z_1\hat{e}_1,z_2\hat{e}_2}) \delta(z_1-z_2)\,,
\ee
where $r_{z_1\hat{e}_1,z_2\hat{e}_2}$ is the comoving distance between these two
points. This is obviously not correct since correlations at different times
or different redshifts exist as cosmological fluctuations are all
related by transfer functions to the initial correlations set by
inflation. Equation~(\ref{WrongLimber}) could naively seem to be a
special case of a Limber approximation, but this is not the case. It is based on the wrong idea that
since we observe on our past light-cone, points observed at very
different redshifts are also necessarily very far apart and thus very
little correlated. But this information is already contained in the
correlation function $\xi_{\rm Gal}(r)$ which dies off for scales
larger than the baryon acoustic oscillations scale, that is for
scales larger than $150$~Mpc.  A first issue arises immediately when
we  consider the correlation (\ref{WrongLimber}) for the same
direction $\hat{e}_1 = \hat{e}_2$, that is with no angular separation. In
that case, the Dirac delta function in Eq. (\ref{WrongLimber}) is equivalent
to changing the shape of the correlation function, causing it to 
vanish for finite distances. It corresponds to white noise in the
redshift dependence which is unphysical. But the most serious issue is
that it leads to the expression of the multipoles given by Eq. (67) of Ref.~\cite{Mairi1}, and this is ill-defined as we shall now show.

\begin{figure}[!htb]
\includegraphics[width=\columnwidth]{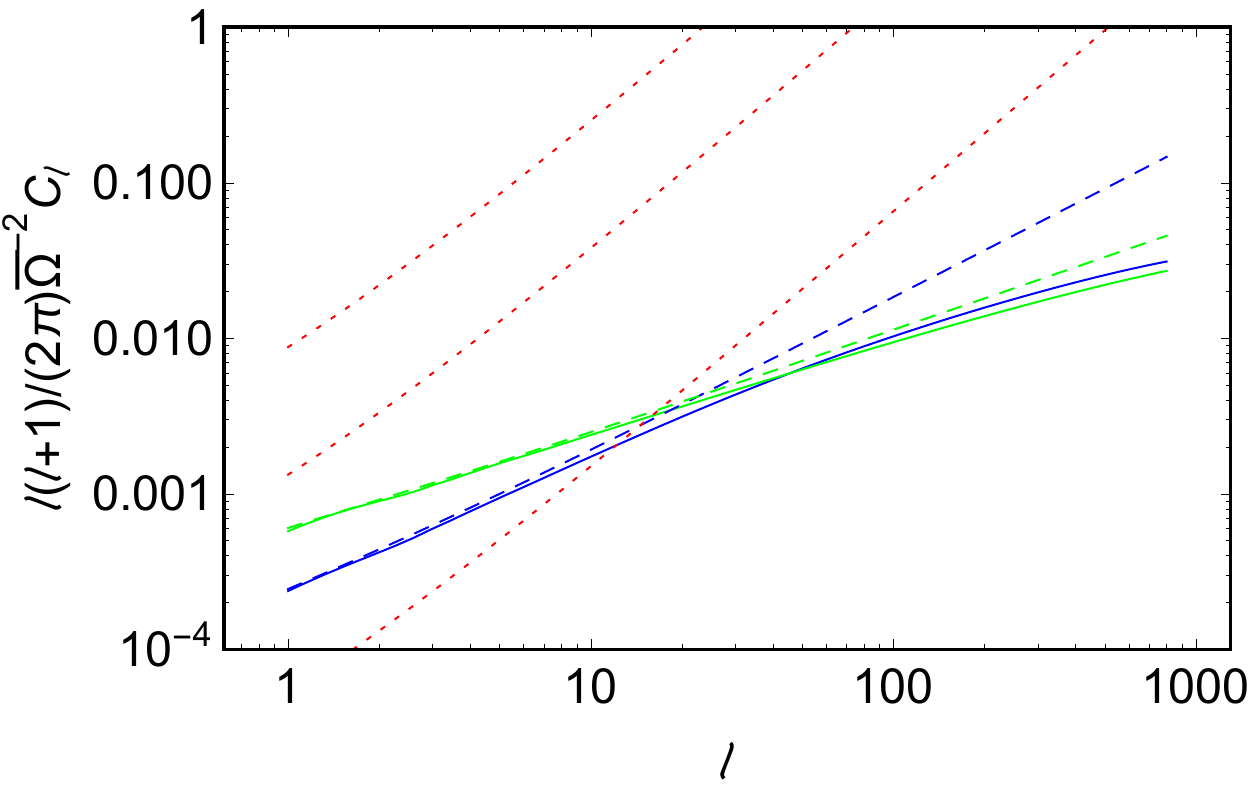}
\caption{Angular power spectrum of relative fluctuations in the
  astrophysical GW background at  $f=32\rm Hz{}$. {\it Red}: multipoles computed
  using both the power-law galaxy spectrum and analytic method of
  Jenkins et al. [Eq.~(\ref{ClMS})
  of this note] with cut-offs $z_{\rm min}=0.0033,0.033,0.33$ from top
  to bottom.  {\it Blue}: multipoles computed using both our galaxy spectrum
  (CMBquick + Halofit) and our standard method to compute the angular power spectrum [essentially
  Eq.~(\ref{ClStandardMethod}) with density growth functions
  properly taken into account], along with our analytic estimate of its
  scaling in dashed line [based on Eq. (\ref{ClProp})]. {\it Green}: multipoles computed using the
  power-law spectrum of Jenkins et al. corrected by
  Eq.~(\ref{DefGrowth}), but using our standard method for the angular power spectrum, along
  with the Limber result (\ref{ExactClLimber}) in dashed line.  In all curves we use our astrophysical model. The blue and green
 curves differ on large angular scales (small $\ell$) because the galaxy power spectra used in the two predictions also differ on large scales (small
 Fourier modes). However, only our galaxy spectrum provides an accurate description of these large
 scales, as shown in the text. The continuous and dashed green curve do not match for large
 $\ell$, even though the Limber approximation
 converges as $1/\ell^2$, because we are using a cut-off $k_{\rm max} \simeq
 5 \,{\rm Mpc}^{-1}$ when evaluating Eq.~(\ref{ClStandardMethod}). Our
result (blue curve) is much less sensitive to this cut-off
because for larger modes we have less power than the power law
spectrum. The blue continuous and dashed curves differ for
large $\ell$ essentially because we also ignored the variations of the densities and $\partial_r \bar \Omega$
when deriving the approximation~(\ref{ClProp}).  The red curves are all incorrect
 because they are based on an ill-defined method to compute the
 angular power spectrum of the background, as explained in the
 text. }\label{ComparisonSpectraPlot}
\end{figure}

In full generality, the monopole of the background\footnote{In this
  note $\bar \Omega$ is defined as the average background per unit of
  solid angle as in Jenkins et al., and it is thus $4\pi$ lower than the $\bar \Omega$ defined in Refs.~\cite{Cusin:2017fwz,Cusin:2018rsq}.} takes the form
\be
\bar \Omega(f) = \int_{0}^{z_{\rm max}} \dd z \partial_z \bar \Omega(f,z)\,,
\ee
meaning that any model boils down to deriving what is the contribution per unit of redshift to the total background.  For instance in Refs.~\cite{Cusin:2017fwz,Cusin:2018rsq}, this is given by
\be
\partial_z \bar \Omega(f,z) \equiv \frac{f}{4\pi \rho_c H(z)} {\cal A}(f,\eta(z))\,.
\ee
Using Eq.~(\ref{WrongLimber}), it is shown in Ref.~\cite{Mairi1} that the multipoles of the anisotropies for the GW background take the form
\be\label{ClMS}
C^{\text{Jenkins}}_\ell(f) = 2 \pi \bar\Omega^{2}(f) {\cal A}_{\rm GW}(f)\, c_\ell
\ee
with\footnote{Beware that ${\cal A}$ of
  Refs.~\cite{Cusin:2017fwz,Cusin:2018rsq} is completely different
  from  ${\cal A}_{\rm GW}$ of Jenkins et al.}
\be\label{fell}
c_\ell \equiv \int_{-1}^{+1}  \dd \cos \theta P_\ell (\cos \theta)
\left[\tan(\theta/2)\right]^{-\gamma}\,,
\ee
\be\label{DefAgw}
 {\cal A}_{\rm GW}(f) \equiv  \bar\Omega^{-2}(f)\int_0^{z_{\rm max}}
 \dd z\frac{d_1^\gamma }{[2 r(z)]^\gamma}\left[\partial_z \bar\Omega(f,z)\right]^2\,.
\ee
We observe that  for $\gamma>1$ the integrand function diverges at $z=0$, unless $\partial_z \bar
\Omega(f,z=0)=0$. 
However, no model would predict reasonably that $\partial_z \bar
\Omega(f,z)$ vanishes at $z=0$, since this would imply that there is
no gravitational-wave production in our vicinity, i.e. no merging events
nearby, which would be in contradiction with the LIGO BH-merger
detections (with luminosity distance roughly in the range 200 to
1500~Mpc, i.e. with redshift in the range 0.05 to 0.3). Given that $z
\simeq H r$ at low redshifts, we see immediately that for $\gamma>1$,
the expression (\ref{DefAgw}) is not defined. The only possibility to
obtain a finite result is to start the integration at a redshift
$z_{\rm min}$ corresponding to a $r_{\rm min}$. But then the result
obtained is completely arbitrary as it strongly depends on the
cut-off. Using our astrophysical model, we infer that to obtain ${\cal
  A}_{\rm GW} $ of the order of $10^{-4}$ as quoted in Jenkins et al.,
one must use $z_{\rm min} \simeq 0.033$ corresponding to a minimum distance $r_{\rm min} \simeq 150 {\rm Mpc}$, which is impossible to justify physically. 
In Fig.~\ref{ComparisonSpectraPlot}, we report the angular power spectrum (\ref{ClMS}) for various cut-off redshifts $z_{\rm min}$. It can be checked that they follow the scaling $c_\ell \propto
\ell^{\gamma-2}$, hence $\ell(\ell+1)C_\ell/(2\pi) \propto
\ell^\gamma$.  Hence, Eq.~(\ref{DefAgw}) does not allow one to obtain the amplitude of the signal. Moreover, as we will show in the following, Eq.~(\ref{fell}) provides the wrong scaling of the angular power spectrum. 

In order to compute correctly the multipoles, and as for any
cosmological observable, one needs to use the expressions for the
angular correlations which use the power spectrum instead of the
correlation function, and which are presented in
Ref.~\cite{Cusin:2017fwz}. When considering only the dominant
effect of galaxy density fluctuations, and estimating these
fluctuations by their value today (hence neglecting the linear
growth), the angular correlation of the gravitational waves background
takes the simple form (omitting the dependence on the GW frequency $f$)
\be
C(\hat{e}_1,\hat{e}_2) \simeq \int \dd r_1 \dd r_2 \partial_{r_1} \bar
\Omega(r_1) \partial_{r_2} \bar \Omega(r_2) \xi_{\rm Gal}(r_{r_1\hat{e}_1,r_2\hat{e}_2})\,,
\ee
where  $\partial_r \bar \Omega(r) = H \partial_z \bar \Omega(z(r))$. Using the transformation~(\ref{HankelPtoxi}) and Eq. (A.12) of
Ref.~\cite{Reimberg2015} we find the classic expression for the corresponding multipoles
\be\label{ClStandardMethod}
C_{\ell}^{\text{std}}\simeq \frac{2}{\pi} \int k^2 \dd k P_{\rm Gal}(k) \left|\int
  \partial_r \bar \Omega(r)j_\ell(k r)\dd r\right|^2\,.
\ee
We refer to this expression for the angular power
  spectrum as \emph{standard} since it is based on the usual
  computational method of angular power spectra.

In practice, it then allows one to use the Limber
approximation~\cite{LoVerde:2008re,PitrouFlat}. One method consists in noticing
that for a test function $f(x)$
\be\label{Limberjl}
\int \dd x j_\ell(x) f(x) \simeq \sqrt{\frac{\pi}{2\ell+1}}f(\ell+1/2)\,.
\ee 
This leads to the two equivalent formulations of the Limber
approximation
\bea
C^{\rm Limber}_\ell &\simeq & \int \frac{\dd r}{r^2} P_{\rm Gal}(k) \left|\partial_r \bar  \Omega (r)\right|^2\,,\label{Limberr}\\
C^{\rm Limber}_\ell & \simeq & \left(\ell+\tfrac{1}{2}\right)^{-1}\int \dd k P_{\rm Gal}(k) \left|\partial_r \bar  \Omega (r)\right|^2\,,\label{Limberk}
\eea
where $k$ and $r$ must satisfy the Limber constraint
\be\label{LimberConstraint}
k\, r = \ell +\frac{1}{2}\,\,.
\ee
In fact, the Limber approximation can also
  be understood as a special case of the flat-sky approximation, and
  in that case a Dirac delta function enforcing equal conformal
  distance correlation appears naturally (see e.g. Ref.~\cite{PitrouFlat},
  section IIIA). 
This is another  way to be convinced that Eq.~(\ref{WrongLimber}) is
an incorrect shortcut.\footnote{From Eq.~(7) of
  Ref.~\cite{PitrouFlat}, and assuming that $P(k) \simeq P(k_\perp)$
  to enforce the Limber approximation, then by replacing Eq.~(\ref{HankelxitoP}), and using that $\int j_0(x)J_0(ax) x
  \dd x$ vanishes for $a>1$ and evaluates to $1/\sqrt{1-a^2}$ for
  $a<1$, we can deduce what the correct Ansatz for the rhs of
  Eq.~(\ref{WrongLimber}) should be. It takes the form $d \widetilde{\xi} (d) \delta(r_1-r_2)$,
where $d \simeq r_1 \theta $ is the comoving separation on the flat sky, and
with the projected correlation $\widetilde{\xi} (d) \equiv 2
\int_0^{\pi/2} (\sin\alpha)^{-2} \xi(d/\sin \alpha)  \dd \alpha$. In particular for
the power law correlation function~(\ref{Defxi}), $\widetilde{\xi}(d) =
\sqrt{\pi}\xi(d)\Gamma[(\gamma-1)/2]/\Gamma(\gamma/2)$. It can then be
checked that the Limber approximation multipoles  (\ref{Limberr})  are
recovered by using the methodology of Ref.~\cite{PitrouFlat} and enforcing the appearance of
$P_{\rm Gal}(k) = P_{\rm Gal}[(\ell+1/2)/r]$ from its power law expression~(\ref{PGalk}).} 

We observe that if the galaxy power spectrum has a power-law functional dependence, the Limber approximation automatically gives the slope of the background angular power spectrum. However, if the functional dependence is more complex, it is useful to introduce the further assumption that the emission is constant in redshift. More in detail, using Eq. (\ref{Limberk}) and assuming further that we can ignore the variations of $\partial_r \bar
\Omega$, we then find that the multipoles are approximately given by
\be\label{ClProp}
C^{\text{Limber+static}}_\ell \propto
\frac{\left(\partial_r \bar \Omega|_{r=0}\right)^2}{\ell+\tfrac{1}{2}} \int_{k_{\rm min}(\ell)} P_{\rm Gal}(k) \dd k\,.
\ee 
where $k_{\rm min}(\ell)$ is set by the fact that there is a maximum distance $r_{\rm max}$ at which we can find GW sources and thus a
minimum Fourier mode set by the Limber constraint~(\ref{LimberConstraint}).
We refer to the expression for the angular power spectrum (\ref{ClProp}) as \emph{Limber+static} since it is based on the
standard computation of the angular power spectrum (\ref{ClStandardMethod}), with the use of the Limber approximation and
simplifying assumptions about the time evolution of sources (which
properly captures the large angle slope of the spectrum). 

We use in the following the Limber approximation to get an estimate of the scaling of the background angular power spectrum, for both the galaxy correlation function of Cusin et al. and Jenkins et al. 
\begin{itemize}
\item  
With our galaxy power spectrum which describes correctly the large
scales, and thus the small Fourier modes, the proportionality relation
(\ref{ClProp}) is insensitive to $k_{\rm min}$ and one can replace it by $0$. Indeed for $k<k_{\rm eq}$ (with $k_{\rm eq} \simeq 0.01 \,{\rm Mpc}^{-1}$ the Fourier mode entering the horizon at matter-radiation equivalence), $P_{\rm Gal}(k) \propto k^\alpha $ with $\alpha
\simeq 1$. In particular we find that $C_\ell \propto \ell^{-1}$, and hence $\ell(\ell+1)C_\ell/(2\pi) \propto \ell$. 

\item 
With the power law spectrum of Jenkins et al. we have $P_{\rm
  Gal}(k) \propto k^{\gamma-3}$, meaning that in eq. (\ref{ClProp}) we are sensitive to
low-$k$ and hence large distance contributions. Fortunately to compute the slope of the angular power spectrum in that case, it is not necessary to assume that variations  $\partial_r \bar
\Omega$ can be neglected, and one can rely on the Limber relation~(\ref{Limberr}) to obtain
\bea\label{ExactClLimber}
C_\ell^{\rm Limber}&\simeq& \frac{g^2 4 \pi
  \Gamma(2-\gamma)\sin(\pi\gamma/2)}{\left(\ell+\tfrac{1}{2}\right)^{3-\gamma}}\nonumber\\
&\times&\int
r \dd r\left( \frac{d_1}{r}\right)^\gamma |\partial_r \bar \Omega(r)|^2\,.
\eea
This is the correct version of Eq.~(\ref{ClMS}), that is of Eq.~(67) in Ref.~\cite{Mairi1}.
Hence we find $C_\ell \propto \ell^{\gamma-3}$, that is $\ell(\ell+1)C_\ell/(2\pi) \propto \ell^{\gamma-1}$.
\end{itemize}

In Fig.~\ref{ComparisonSpectraPlot} (blue and green lines) we compare
the multipoles as obtained with our power spectrum to the ones obtained using the same astrophysical
model, but with the power spectrum of Jenkins et al. corrected by the
growth factor~(\ref{DefGrowth}).  Both these curves are obtained using
the standard (correct) method to derive the angular power spectrum of
the background, see Eq. (\ref{ClStandardMethod}).  For our spectrum we also show in dashed blue line the approximate slope based on
Eq.~(\ref{ClProp}) whereas for the power law spectrum of Jenkins et al. we show in
dashed green line Eq.~(\ref{ExactClLimber}). It can be checked visually that the power-law spectrum of Jenkins et al. overestimates the low
$\ell$ multipoles as they benefit the most from the low $k$ Fourier
modes which are incorrectly described by the power law. 
We note that the fact that the analytic approach of
  Jenkins et al., which we have shown to be inconsistent, is in
  agreement with their result obtained using a galaxy catalogue extracted from the Millennium simulation, is rather puzzling 
   and calls for an explanation. \\

To conclude in a less technical way, while we
agree that the discrepancy between the predictions of the
astrophysical GW angular power spectrum of Refs.~\cite{Cusin:2018rsq}
and Jenkins et al. mostly arises from the description of the clustering, we
argue that the treatment of Jenkins et al. is flawed since their galaxy correlation
function does not give a realistic description of large scales and
furthermore their analytic treatment of the background correlation function is mathematically wrong. This
emphasizes that a precise prediction of the stochastic GW angular
spectrum requires both a proper astrophysical description of the BH
formation and distribution but also of the cosmology and the
distribution of the large scale structure. Fortunately, the latter is
well-understood theoretically and under control from an observational
point of view. In the above discussion we have detailed the agreement of our
description~\cite{Cusin:2018rsq} with galaxy power spectra
observations and the reasons why the one of Jenkins et al. is unsatisfactory. Further work should now be done on elucidating the dependence of the  GW stochastic background anisotropies on astrophysical models and the possibility to constrain them.\\

\noindent{\bf Note added on CMBquick}. Let us take the opportunity to emphasize that the code CMBquick~\cite{CMBquick} is distributed under the General Public License whose disclaimer, actually recalled in the code, states that it is {\it without warranty of any kind  [...] including the implied warranties [...] of fitness for a  particular purpose. The entire risk as to the quality and performance of the program is with you}. CMBquick is only meant as a pedagogical tool for CMB correlations computations, and more generally linear transfer functions, and as such it is provided with default parameters which are only suited for CMB computations. Furthermore it is written without any documentation about the astrophysical background computations and the conventions of normalisation. It is therefore very likely that the numerical results obtained from CMBquick in Ref.~\cite{Mairi2} are also not precise.


\bibliography{myrefs}

\end{document}